\input epsf
\documentstyle[twoside,fleqn,espcrc2]{article}
\font\tencop=bbm10
\font\sevencop=bbm7 \font\fivecop=bbm5
\newfam\copfam\scriptscriptfont
\copfam=\fivecop\textfont\copfam=\tencop\scriptfont\copfam=\sevencop

\newcommand{\Z}{{\bf Z}}
\newcommand{\IM}{\mathop{\Im m}\nolimits}
\title{Duality and BPS spectra in N=2 supersymmetric QCD}

\author{Frank Ferrari
\address{Laboratoire de Physique th\'eorique de l'\'Ecole
Normale Sup\'erieure\\
Unit\'e Propre de Recherche 701 du CNRS\\
24, rue Lhomond, 75231 Paris Cedex 05}}

\begin{document}

\begin{abstract}
I review, with some pedagogy,
two different approaches to the computation of
BPS spectra in $N=2$ supersymmetric QCD with gauge group SU(2).
The first one is semiclassical and has been widely used in the
literature. The second one makes use of constraints coming from
the non perturbative, global structure of the Coulomb branch
of these theories. The second method allows for a description
of discontinuities in the BPS spectra at strong coupling, and
should lead to accurate tests of duality conjectures in $N=2$
theories. 
\end{abstract}

\maketitle

\setbox1=\hbox{LPTENS-96/52}
\setbox2=\hbox{\tt hep-th@xxx/9611012}
\makeatletter
\global\@specialpagefalse
\def\@oddhead{\ifnum\c@page>1\reset@font{\sl\rightmark}\hfil \rm\thepage\fi
\ifnum\c@page=1
\hfill\vbox{\box1\box2}\fi }
\def\@evenhead{\reset@font\rm \thepage\hfil\sl\leftmark}

\section{General overview}
The understanding of
strong-weak coupling dualities, or S dualities, in supersymmetric
gauge theories (on which I will focus)
and superstring theory, is certainly the most
outstanding problem in high energy physics at present. Typically,
S duality relates a theory A with coupling constant $g$ to another
theory B with coupling constant $1/g$. Testing
a S duality conjecture thus requires the knowledge of some non perturbative
informations, at least for one of the theory. In some cases,
one expects that A and B coincide, \hbox{i.e.} they have the same lagrangian,
with different parameters. Such ``self-dual'' field theories are
certainly the most fascinating one from the theoretical point of
view. The $N=4$ theory is strongly believed to belong to this class,
as well as other, more interesting and more difficult to study,
$N=2$ theories.
 In other cases, the duality is more subtle, and can only
be valid in the low energy limit, as in the asymptotically free
theories.

A complete proof of the duality properties in these theories seems out
of reach at present. However, some accurate tests can be done.
One of them is to look at the low energy effective actions of the
theories conjectured to be dual, and check whether a sensible
duality transformation can be found relating them.
This approach strongly suggests that the $N=2$
theory obtained from the $N=4$ theory by adding a term
$m\,{\rm tr}\, \Phi ^2$ to the superpotential, as well as the
$N=2$ theory with four flavours of quarks in the fundamental
representation of the gauge group (for SU(2)), should be
self-dual \cite{SWI,SWII}. 

Another test is to look at the BPS spectra. This test is more
accurate than simply looking at the low energy effective action,
since it directly probes the Hilbert space of states. When the
spectrum is stable, one can deduce strong coupling results from
weak coupling, semiclassical, calculations. For instance,
this was done in
\cite{sen,por,segal} for the SU(2) $N=4$ theory, with results 
in agreement with S duality, as
we will see below. However, when the number of supersymmetries decreases
down to 2, there may exist curves in moduli space, which generically
separate a strong coupling 
from a weak coupling region, across which the BPS spectrum is
discontinuous \cite{SWI,SWII}. In these cases, the semiclassical
method is hopeless, since it is only valid at weak coupling.
One must devise another method to
understand the jumping phenomenon,
then compute the strong coupling spectrum and check if it is 
compatible with S duality. Note that in the $N=4$ theory, complete
SL$(2,\Z )$ invariance requires that all the BPS states $(n_e,n_m)$,
where $n_e$ and $n_m$ are relatively prime integers 
corresponding respectively 
to the electric and magnetic charges, must exist as quantum stable
states in the theory \cite{sen}. This is no longer the case when 
some discontinuity curves are present: if a duality transformation
relates two region in the parameter space which are separated by such 
a curve, we can only say that the spectrum in one region must be the
dual of the spectrum in the other region.

In this short lecture, I will describe how the strong coupling
BPS spectra have been computed in the asymptotically free theories, 
following \cite{FB,BF}. These theories are not self-dual,
though electric-magnetic duality still plays a profound r\^ ole
\cite{SWI,SWII}. They provide an example where the ``duality'' group
of the low energy effective action (or of the spectral curve associated
with it), typically a subgroup of finite index of SL$(2,\Z )$,
has nothing to do with the duality transformations which can be thought
as being valid quantum mechanically. This provides the first steps
toward the study of the expected self-dual field theories.
I also give a short introduction to the semiclassical quantization,
which in any case provides useful informations, in particular 
concerning the quantum numbers carried by the solitonic states.
\section{The semiclassical quantization}
In the bosonic sector, the classical study of the BPS 
monopole configurations amounts
to solving the Bogomol'nyi equation \cite{BOG}
\begin{equation}
\label{bog}
B=\pm {\rm D}\phi, 
\end{equation}
where
$B$ is the (non abelian) magnetic field and $\phi $ the Higgs
scalar transforming in the adjoint representation of the gauge group
SU(2).
The configuration of $\phi $ at infinity gives an element of
$\pi _2(S^2)=\Z$, which is the magnetic charge $n_m$. At fixed $n_m$,
the set of solutions of (\ref{bog}) 
can be parametrized by $4n_m$ real parameters. For $n_m=1$, these
are the position of the center of mass of the soliton, as well as
a fourth periodic collective coordinate corresponding to the electric
charge. For $n_m$ widely separated monopoles of charge $1$, which correspond
to a configuration of global charge $n_m$, we will have $4$ parameters
describing the global motion and electric charge, and $4n_m-4$
parameters describing the relative motions and electric charges.
The $4n_m$ dimensional parameter (moduli) space ${\cal M}_{n_m}$
was studied in great detail in \cite{ATH}, and has nice
mathematical properties. In particular, it
is a hyperk\" ahler manifold, as we will
see below. The knowledge of the explicit form of the metric on 
${\cal M}_{n_m}$ allows one to study the 
classical low energy dynamics of a $n_m$-monopole configuration,
which corresponds to geodesic motion \cite{manton}.
This can be done in full generality when $n_m=2$, which is the
only case where the metric is known \cite{ATH} (see \cite{sut}
and references therein for recent developments).

However, the study of the classical motion is not enough for our
purposes. What we need is to investigate the {\em quantum} stability of
multi-monopole BPS configurations. This requires to look at
the quantum mechanics associated with the classical
dynamical system describing the geodesic motions, and look for stable
bound states satisfying the Bogomol'nyi bound.
Since $\dim {\cal M}_{n_m}=4n_m$, this quantum mechanics has
$4n_m$ bosonic degrees of freedom $z^\alpha $. Because of supersymmetry,
one would expect to have in addition $4n_m$ fermionic collective
coordinates $\lambda ^\alpha $. 
These indeed come from the zero modes of the Dirac
equation associated with the adjoint fermions, whose number can
be computed using Callias' index theorem \cite{CAL}.
The action then reads \cite{gau1}
\begin{equation}
\label{ac1}
S=\int {\rm d}t\,G_{\alpha\beta }(z)(\dot z^\alpha \dot z^\beta
+i\lambda ^\alpha D_t \lambda ^\beta).
\end{equation}
$G_{\alpha\beta }$ is the metric on ${\cal M}_{n_m}$, $D_t$ the covariant
derivative associated with it, and $\dot z={\rm d}z/{\rm d}t$.
As the original theory has two supersymmetries in four 
space-time dimensions,
which correspond to eight real supersymmetry generators, and as
a BPS configuration breaks half of the supersymmetries, $S$ must have
four real supersymmetries. This is possible if and only if
the target space ${\cal M}_{n_m}$ is hyperk\" ahler \cite{AG}. Note that this 
nice mathematical property of ${\cal M}_{n_m}$ can 
be proved independently \cite{ATH}.
If $N_f$ matter hypermultiplets are also present (we will limit ourselves to
zero bare masses), we will have $2n_mN_f$ additional fermionic
zero modes $\kappa ^{jA}$, $1\leq j\leq 2N_f$, $1\leq A\leq n_m$.
The action will then pick up a new term \cite{ceder,gau2}
\begin{equation}
\label{ac2}
S_{\rm m}=\int {\rm d}t\, \Big( i\kappa ^{jA}{\cal D}_t \kappa ^{jA}
+{1\over 2}F^{AB}_{\alpha\beta }\lambda ^\alpha \lambda ^\beta
\kappa ^{jA}\kappa ^{jB}\Big),
\end{equation}
where ${\cal D}_t$ is the covariant derivative corresponding to a
natural O($n_m$) connexion related to the isospinor fermionic
zero modes, and $F$ the associated curvature two-form. The total action
still has four real supersymmetries in spite of the mismatch
between the number of fermionic and bosonic zero modes,
supersymmetry being non linearly realized. 
The standard quantization procedure leads to 
\begin{equation}
\label{algebra}
\{ \lambda ^\alpha ,\lambda ^\beta \} = \delta ^{\alpha\beta };\ 
\{ \kappa ^{jA} ,\kappa ^{lB} \} = \delta ^{jl} \delta ^{AB},
\end{equation}
which are Clifford algebras whose representation theory is well known.
The hamiltonian $H$
associated with the action $S+S_{\rm m}$
is the square of a Dirac operator coupled to the O($n_m$) connexion.
Its normalizable zero modes will correspond to quantum mechanically
stable BPS states (note that
only the zero modes will correspond to states saturating the
Bogomol'nyi bound). Finding these zero modes
is a very hard mathematical problem
which requires the knowledge of the metric on ${\cal M}_{n_m}$
and the use of advanced index theory. This is why only
very partial results have been obtained up to now in this direction
\cite{gau2}. 
There exists however a trick, introduced and applied with some success
by Porrati in \cite{por} for the $N=4$ theory. Since $H$ is a supersymmetric
hamiltonian, the existence of zero modes is equivalent to the fact that
supersymmetry is not broken in the supersymmetric quantum mechanics.
A very convenient way of proving that susy is not broken is then to
compute the Witten index \cite{witin}. 
The main drawback of this method is that it does not allow
to count the number of the zero modes, and thus leads only to a partial
computation of the BPS spectrum. It also
requires the use of not so well established results concerning the
asymptotic behaviour of the multi-monopole moduli space
${\cal M}_{n_m}$.

We will not pursue this route here. 
But before presenting a completely different method in the next Section,
let us briefly discuss the quantum numbers 
carried by the BPS states. From the representation theory of the
algebras (\ref{algebra}) we know that the wave function of any BPS state
can be written
\begin{equation}
\label{wf}
\vert\Psi\rangle =f(z^\alpha )\,\vert\psi _0 \rangle\otimes
\vert\psi _s \rangle\otimes\vert\psi _{\rm f}\rangle .
\end{equation}
$f(z^\alpha )$ is the bosonic part, which does not carry spin nor
flavour indices. 
$\vert\psi _0 \rangle $ comes from the fermionic collective coordinates
associated with the center of mass motion. This part of the wave function
carries spin, as isovector zero modes $\lambda $ do, 
and puts the BPS state into a
$N=2$ multiplet. $\vert\psi _s \rangle $ also carries spin, and
exists only when $n_m\geq 2$ since it corresponds to fermionic coordinates
associated with the relative motion. $\vert\psi _{\rm f}\rangle $
is due to the isospinor zero modes $\kappa $, which do not carry spin but
put the states into flavour multiplets. For instance, it is clear from
(\ref{algebra}) that for $n_m=1$ the BPS states are in spinorial
representation of the flavour symmetry group Spin($2N_f$).
In general, one can obtain constraints
relating the flavour representations and
the electric charge carried by the states 
using the form (\ref{wf}) of the wave function and the fact that
the isospinor zero modes pick up a minus sign under a
$2\pi $ electric rotation. These constraints were listed \hbox{e.g.} in 
\cite{BF};
they limit the possible decay reactions across curves
of marginal stability and thus provide consistency checks of the
predicted strong and weak coupling spectra \cite{BF}.

Let me close this discussion of the quantum numbers with a remark.
In the $N=2$ theory, there are only two complex zero modes carrying
spin
for $n_m=1$. These are the spin $1/2$ isovector zero modes. 
We thus see that a monopole of unit magnetic
charge cannot carry spin greater that $1/2$ (they lie in standard
$N=2$ matter hypermultiplet) and by the way cannot be the dual of the
W bosons (in the $N=4$ theory there are twice as many isovector zero modes
and this problem disappears). However, for $n_m\geq 2$, we have
additional zero modes carrying spin, which are included in 
$\vert\psi _s\rangle $,
and thus these states may be dual to the W. This is what one expects
to occur in
the $N=2$ theory with four flavours.

\section{The non perturbative approach}
Now I wish to present the method used in \cite{FB,BF}, where
the BPS spectra of the asymptotically
free theories ($0\leq N_f\leq 3$) were rigorously computed,
both at weak and strong coupling.
This leads in particular to the first description of the decay reactions
across the curves of marginal stability at strong coupling (an
alternative approach from string theory has appeared since then 
\cite{ler}).
The method is non perturbative in nature and uses constraints coming
from the global analytic structure of the Coulomb branch of the theories.
The remarkable, and unexpected, result that emerges is the following:
there is one and only one BPS spectrum compatible with the global low
energy structure of the theories.
Whether this is a very general statement, or is limited to the
special cases studied so far, is an open question.
\subsection{General analysis}
In the theories under study, the scalar potential has flat directions
which cannot be lifted quantum mechanically due to tight constraints
coming from supersymmetry \cite{sei}. These flat directions
generate a moduli space which has a Coulomb branch along which
the gauge group SU(2) is spontaneously broken down to U(1) by the
Higgs expectation value $\langle\phi\rangle =a\sigma _3$. A good,
gauge invariant coordinate along the Coulomb branch is
$u=\langle {\rm tr}\,\phi ^2 \rangle$. The low energy, wilsonian,
effective action can be expressed in terms of a single
holomorphic function $\cal F$ (the prepotential) because of $N=2$
supersymmetry as
\begin{eqnarray}
\label{leff}
\lefteqn{{\cal L}_{\rm eff}=
{1\over 8\pi}\,\IM\,\Bigl[ 2\int d^2\theta d^2\overline{\theta}
\, {\cal F}'(A)
\overline{A}+}\nonumber \\
&&\qquad\qquad\qquad\qquad {\int d^2\theta\, {\cal F}''(A)W^2\Bigr] .}
\end{eqnarray}
$(W,A)$ is the $N=2$ abelian massless vector multiplet. One introduces
traditionally $a_D={1\over 2}{\cal F}'(a)$ and the coupling constant
is then $\tau (a)={\rm d}a_D/{\rm d}a=\theta /\pi + {\rm i}
8\pi /g^2$, $\theta $ representing the low energy $\theta $ angle and
$g$ the gauge coupling constant.

For concreteness, I will exclusively study the $N_f=1$ theory in the
following. It exhibits all the main features of the other asymptotically
free theories. Asymptotic freedom is used here to deduce the 
form of the gauge coupling $g(a)$ when $a$ goes to
infinity from the perturbative $\beta $
function $\beta =-3g^3/(16\pi ^2)$: 
\begin{equation}
{1\over g^2} ={3\over 8\pi ^2}\,\ln{\vert a\vert\over\Lambda } .
\end{equation}
Holomorphy, and the fact that $a(u)\sim\sqrt{u/2}$, then yield
\begin{equation}
a_D(u)\sim {3{\rm i}\over 2\pi }\,\sqrt{u/2}\,\ln {u\over\Lambda ^2}.
\end{equation}
$\Lambda \propto a\exp (-8\pi ^2/3g^2)$ 
is the dynamically generated scale of the theory. As we will see
below, it has a clear physical signification since it gives the
typical scale at which singularities appear on the Coulomb branch.
The $n$ instanton corrections to the
perturbative results are proportional to $\exp
(-8\pi ^2n/g^2)\propto\Lambda ^{3n}$.
Because of a flavour parity symmetry existing in the theory,
only even numbers of instantons contribute
(see \hbox{e.g.} \cite{SWII}) and we expect to have
\begin{eqnarray}
\label{asymp}
\lefteqn{a(u)=\sqrt{u/2}\left[ 1+\sum _{k=1} ^\infty a_k
\left( {\Lambda ^2\over u}\right)^{3k} \right] }\nonumber \\
\lefteqn{a_D(u)={3{\rm i}\over 2\pi }\sqrt{u/2}\ln {u\over\Lambda ^2} + 
\sqrt{u}\sum _{k=1}^{\infty} {a_D}_k
\left( {\Lambda ^2\over u}\right)^{3k}\!\!\!\!\! . } 
\end{eqnarray} 
This is a non perturbative, but still semiclassical, formula, since
it comes from the application of the steepest descent method to the
functional integral.
In particular, the series have a finite radius of convergence, and
we must find an analytic continuation in order to have a global
description.  
How to find this analytic continuation was done by
Seiberg and Witten in their celebrated papers \cite{SWI,SWII}.
\subsection{Results from Seiberg and Witten and general remarks}
Using some physical arguments, it was shown in \cite{SWII} that
the low energy effective action 
of the $N_f=1$ theory under study
has three singularities at strong
coupling, due to dyons of unit magnetic charge becoming massless.
Probably the most important lesson of \cite{SWI,SWII} was
to explain how to compute the asymptotics of
$a$ and $a_D$ near such strong-coupling singularities.
The idea consists in using electric-magnetic duality rotations
to couple locally the soliton becoming massless with the 
$N=2$ abelian vector multiplet contained in ${\cal L}_{\rm eff}$.
One then uses the fact that an abelian gauge theory is weakly
coupled in the infrared to deduce the asymptotics of $g$
using the one-loop $\beta $ function of the low energy theory.
This can also be done directly in the microscopic theory, as was
pointed out recently by the present author \cite{ferrari}. Using these
asymptotics, the monodromy matrices around the singular points can
be deduced. For one hypermultiplet $(n_e,n_m)$ becoming massless,
it is in the conventions of \cite{FB,BF}
\begin{equation}
\label{monod}
M_{(n_e ,n_m)}=\pmatrix{1-n_e n_m & n_e^2 \cr
-n_m^2 &1+ n_e n_m\cr }.
\end{equation}
One important point to realize is that the structure of the monodromy
group is extremely constrained by the discrete 
$\Z _3$ symmetry acting on the Coulomb branch. This discrete symmetry
comes from the anomaly free part $\Z _{12}$
of the U(1)$_{\rm R}$ symmetry of the
classical theory, under which $u$ has charge 4. It imposes that
the three singularities must be at the vertices of an equilateral
triangle, say at $u_1=e^{{\rm i}\pi /3}$, $u_2=-1$ and 
$u_3=e^{{-\rm i}\pi /3}$, choosing $\Lambda\sim 1$. 
Moreover, if $(n_e,n_m)$ becomes massless
at $u_3$, then $(n_e-n_m,n_m)$ will be massless at $u_1$
and $(n_e-2n_m,n_m)$ at $u_2$. As noticed in \cite{BF}, consistency
with the monodromy at infinity
\begin{equation}
M_{\infty}=\pmatrix{-1 &\hfill 3 \cr\hfill 0 &-1\cr}
\end{equation}
actually implies that $n_m=\pm 1$. We will choose to have 
$(0,1)$ massless at $u=u_3$.

To obtain an explicit solution for $a_D$ and $a$ is now a purely
mathematical exercise. Using the theory of differential equations,
it is not so difficult to find a
second order linear differential equation whose solutions have the
required monodromies. Nevertheless, the most elegant and general way of doing
this \cite{SWI,SWII}, keeping in mind that the monodromies are elements
of SL$(2,\Z)$, is to find a family of cubic complex curves (tori)
parametrized by $u$, whose modular parameter is determined
modulo the modular group at fixed $u$. $a$ and $a_D$ are then expressed
as period integrals, and the differential equations are nothing but
the Picard-Fuchs equations associated with them. Then one has
to solve these differential equations (or compute directly the
periods integrals) to obtain the solution explicitly. The latter allows
to obtain the curve of marginal stability, and exhibits the global
analytic structure, two crucial ingredients for our purposes. 
\subsection{Analysis of the explicit solution}
The solution, found in \cite{BF}, reads for $-2\pi /3\leq\arg u\leq 0$ 
\begin{eqnarray}
\label{solution}
a(u)&=&\sqrt{u/2}\, 
F\left( -{1\over 6}\raise 2pt\hbox{,} 
{1\over 6}\raise 2pt\hbox{,}1;-{1\over u^3}\right)\nonumber \\
a_D(u)&=&e^{-2{\rm i}\pi /3}\,
{\sqrt{2}\over 12} \left( u^3 +1\right)\nonumber\\
&&\qquad\qquad\quad
F\left( {5\over 6}\raise 2pt\hbox{,} {5\over 6}\raise 2pt\hbox{,}2;
1 + u^3 \right),
\end{eqnarray}
and we have for all $u$
\begin{eqnarray}
\label{sol2}
\pmatrix{ a_D\cr a\cr } \left( e^{\pm 2i\pi/3} u\right)
&=& e^{\pm i\pi/3} G_{W\pm} \pmatrix{ a_D\cr a\cr } (u),\nonumber\\
G_{W\pm} &=& \pmatrix{ 1& \mp 1\cr 0 & \hfill 1\cr}.
\end{eqnarray}
The latter equation reflects the $\Z _3$ symmetry acting on the
Coulomb branch. The mass of any BPS states $(n_e,n_m)$ can then be computed
very explicitly using the formula \cite{SWI,SWII}
\begin{equation}
\label{BPS}
m=\sqrt{2}\,\vert an_e - a_D n_m\vert.
\end{equation}

The curve $\cal C$ of marginal stability across which the spectrum
of BPS states is discontinuous is the locus of the points $u$ where
the two dimensional lattice generated by $a$ and $a_D$ collapses
to a line:
\begin{equation}
{\cal C}=\left\{u\mid\IM {a_D\over a}=0 \right\}\cdotp
\end{equation}
On this curve, states $(n_e,n_m)$ which are ordinarily stable
(\hbox{i.e.} $n_e$ and $n_m$ are relatively prime, or the state corresponds
to a bound state at threshold like the W bosons $(\pm 2,0)$) may
become unstable and ``decay'' \cite{SWI,SWII,FB,BF}.
$\cal C$ was computed numerically in \cite{BF}. It looks like a circle,
and contains the three singularities as is clear from
(\ref{BPS}). It is depicted in the Figure, where the cuts of the
functions $a_D$ and $a$ are also represented. 
\vskip 1cm

\epsfxsize=7cm
\epsfbox{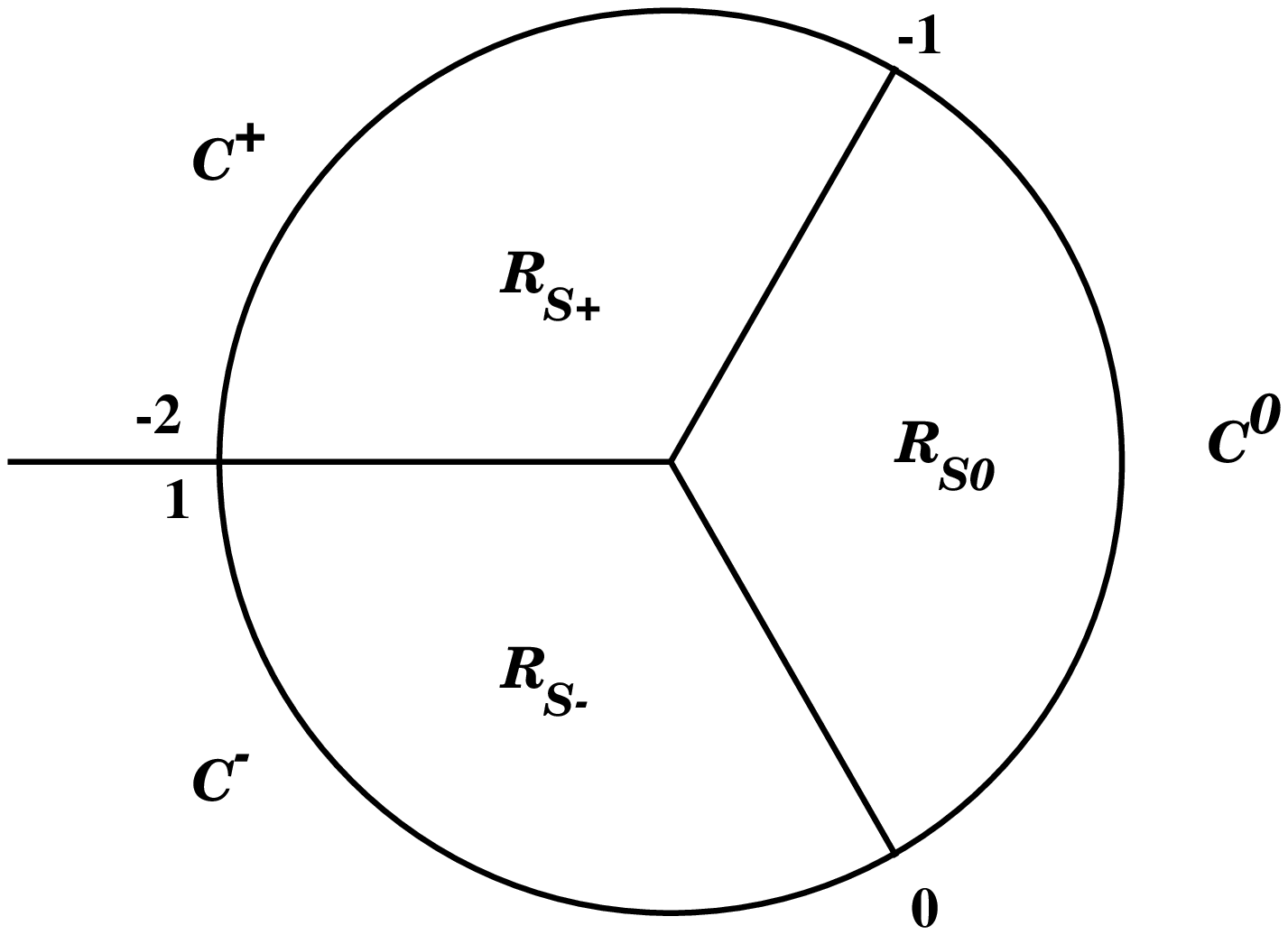}
\vskip .5cm
Figure. The curve of marginal stability ${\cal C}$ passes through
the three cubic roots of -1. It is almost a circle. The numbers 1, 0
-1, -2 indicate the values taken by $a_D/a$ along the curve. The definitions
of the various portions of the curve and of the strong coupling region
are indicated.
\vskip 1cm
The fact that the cuts 
separate the strong coupling region $R_S$ (inside the curve) into three
parts $R_{S+}$, $R_{S-}$ and $R_{S0}$ has a crucial physical meaning.
To understand this, consider a BPS state $(n_e,n_m)$, say in $R_{S0}$.
Its mass is given by (\ref{BPS}). Now, vary $u$ continuously
inside the strong coupling region
and go for instance in the region $R_{S+}$. Physically, nothing
happens when crossing the cut, since the BPS state remains stable
(we do not cross the curve of marginal stability ${\cal C}$ in this process).
In particular the mass of the state must vary continuously, and thus
is given in $R_{S+}$ by $m=\sqrt{2}\,\vert\tilde an_e-\tilde a_D n_m\vert$
where $\tilde a$ and $\tilde a_D$ are the analytic continuations in $R_{S+}$
of $a$ and $a_D$ through the cut separating $R_{S0}$ and $R_{S+}$.
The relation between $(a_D,a)$ and $(\tilde a_D,\tilde a)$ is given
by the monodromy around $u_1=e^{{\rm i}\pi /3}$. 
If we insist in using the solution
$(a_D,a)$ given by (\ref{solution}, \ref{sol2}) all through the $u$-plane,
we see that one cannot label a BPS state by a unique set of
quantum numbers all through the strong coupling region. If
$(n_e,n_m)$ is used in $R_{S0}$, then we must use
$(\tilde n_e,\tilde n_m)$ such that $\vert a\tilde n_e -a_D \tilde n_m\vert
=\vert\tilde an_e-\tilde a_D n_m\vert$ in $R_{S+}$.
This is equivalent to the fact that the SL$(2,\Z)$ bundle over
the Coulomb branch, of which $(a_D,a)$ and $(n_e,n_m)$ are sections,
is not trivial.	The strong coupling monodromies give the transition
functions. Explicitly, denoting by $p$ the locally constant
section representing a given BPS state, we have \cite{BF}
\begin{eqnarray}
\label{trans}
\lefteqn{p\equiv (n_e,n_m)\quad {\rm in}\quad R_{S0}}\nonumber\\
&\qquad\Leftrightarrow & p\equiv \pm(n_e, n_m+n_e)\ {\rm in}
\ R_{S-}\nonumber\\
&\qquad\Leftrightarrow & p\equiv \pm(2n_e+n_m,-n_e)\ {\rm in} 
\ R_{S+}.
\end{eqnarray}
Some remarks concerning these formulas are worthwhile.

First, note that a state becoming massless must exist at
strong coupling, since one can cross
$\cal C$ precisely at the point where it is massless and thus stable
(it is the only massless charged state at this point). For instance,
one can follow continuously the state $(0,1)$ from weak coupling to
strong coupling crossing $\cal C$ at $u=u_3=e^{-{\rm i}\pi /3}$.
Entering in the strong coupling region, one can choose in this case to go
either in $R_{S-}$ or in $R_{S0}$. Since the transition from
weak coupling is continuous, the state $(0,1)$ should be represented 
by the same quantum numbers in $R_{S-}$ and $R_{S0}$,
and these quantum numbers can be computed semiclassically.
Note that this is valid for all the quantum numbers
eventually carried by the state. For the electric and magnetic
charge, this implies in particular that $(0,1)$ must represent
the same state in $R_{S0}$ and in $R_{S-}$, which is indeed the case,
see (\ref{trans}). More generally, this provides a physical interpretation
of the fact that $(n_e,n_m)$ is always an eigenvector of eigenvalue 1
of the monodromy matrix $M_{(n_e,n_m)}$ (\ref{monod}).

Second, note that the transformation rules (\ref{trans})
intimately mix the electric and magnetic quantum numbers.
For instance, the magnetic monopole $(n_e=0,n_m=1)$ becoming massless at
$u=u_3$ will be described by $(n_e=\pm 1,n_m=0)$ in $R_{S+}$!
This phenomenon nicely illustrates the fact that the distinction
between the electric and magnetic quantum numbers is very unclear
at strong coupling, unlike at weak coupling where they have a 
completely different origin.
\subsection{The semiclassical spectrum}
Let us quit for a moment the strongly coupled physics, and focus
on the weak coupling spectrum $S_W$. Next I will present an argument
\cite{FB,BF}
which allows to completely determine it, in a surprisingly easy
way when one has in mind that what we really do is to count zero
modes of a supersymmetric hamiltonian, as explained in Section 1,
or equivalently to compute the cohomology of
very complicated manifolds!

First, the elementary excitations $\pm (1,0)$ (quarks) and
$\pm (2,0)$ (W bosons) must be in $S_W$, as well as the states
responsible for the singularities, that is $\pm (1,1)$,
$\pm (0,1)$ and $\pm (-1,1)$. Looping around the point at
infinity, one can then deduce that all the states
$\pm (n_e,1)$ are in $S_W$, for all integers $n_e$. Note that
this means that electric charge is quantized, which simply reflects that 
the collective coordinate associated with it is a periodic variable.
What I wish to prove now is that no states of magnetic charge
greater than or equal to two exist. Suppose the contrary, and let
$(n_e,n_m)$ be such a state. Using the monodromy at infinity
as above, one deduces immediately that the states $(n_e-3kn_m,n_m)$ 
are also in $S_W$ for all integers $k$. Let us look at the particular state
$(n_e-3k_0n_m,n_m)$ where $k_0$ is chosen such that 
$(n_e-3k_0n_m)/(n_m)\in [-2,1]$. Because $a_D/a$ takes all values
in the interval $[-2,1]$ along the curve of marginal stability
$\cal C$, which follows most easily from the monodromy at infinity,
there exists a point $u^*\in \cal C$ such that
$(n_e-3k_0n_m)a(u^*) - n_m a_D(u^*)=0$. This means that $(n_e-3k_0n_m,n_m)$
is massless at $u^*$, which is impossible as the only singularities
on the Coulomb branch are those due to the states $n_m=1$ mentioned
above. This complete the proof.
\subsection{The strong-coupling spectrum}
We now have all the necessary ingredients to determine
the strong coupling spectrum. We already know that the three
states responsible for the singularities are in $S_S$, as
they are stable across the points where they are massless.
Actually, there cannot be any other state. Note first that
$a_D/a$ varies from 1 to 0 along ${\cal C}^-$, from 0 to $-1$
along ${\cal C}^0$, and from $-1$ to $-2$ along
${\cal C}^+$. Note also that the part ${\cal C}^0$ of the
curve borders the region $R_{S0}$, ${\cal C}^+$ the region
$R_{S+}$, ${\cal C}^-$ the region $R_{S-}$ (see the Figure), 
and that different
quantum numbers must be used to describe a BPS state in these
different regions (\ref{trans}).
Thus, a state represented by $(n_e,n_m)$ in $R_{S0}$, $n_m\not =0$,
is never massless if and only if $r=n_e/n_m\notin [-1,0]$,
$n_e/(n_m+n_e)=r/(1+r)\notin [0,1]$ and $-(2n_e+n_m)/n_e=-2-1/r\notin
[-2,-1]$. But this is impossible from very elementary analysis!
The case $n_m=0$ is left to the reader, and we arrive at our
main conclusion: in the strong coupling region, only the states
responsible for the singularities exist. In particular,
no normalizable quantum states correspond to
the W bosons or to the quarks 
in the theory, though they appear as elementary
fields in the lagrangian! Note that this fact can be suspected
using an independent (heuristic) argument for the W (applied
in \cite{lind} to the pure gauge theory).

I wish to conclude this subsection by pointing out that we do not
use explicitly the global discrete symmetry in the reasoning above.
However, the analytic structure is strongly constrained by this
symmetry, as already noted above, and what we do is strictly equivalent
to the reasoning favoured in \cite{FB,BF}. Using explicitly the
global discrete symmetry would allow to work in a fixed region, say
$R_{S0}$. It also leads to the funny conclusion that the BPS states
must come in multiplets of the symmetry, though it is
spontaneously broken, an aspect emphasized in \cite{BF2}.
\subsection{Fun with the quantum numbers}
In finding two different spectra at weak coupling and
strong coupling, we have predicted a set of decay reactions
across the curve $\cal C$ (for a discussion of what these
``decay reactions'' really are, see \cite{BF}).
All these decay reactions should be compatible with the 
conservation of the quantum
numbers carried by the BPS states: mass, spin, electric, magnetic
and flavour charges. These quantum numbers are unambiguously determined
at weak coupling. For instance, the flavour charge $F$ (which is
an abelian U$(1)$ charge for $N_f=1$) is $+1/2$ for
a state $(2k,1)$ and $-1/2$ for $(2k+1,1)$.
At strong coupling, there is a sign ambiguity in $n_e$, $n_m$
and $F$ because the transition functions (\ref{trans}) are
only determined up to a sign. It can be shown that this sign ambiguity is
lifted if one wants the decay reactions to be possible. I refer
the reader to  \cite{BF} for more details.
\section{Conclusion}
We have presented two methods which allow to compute the BPS
spectra in $N=2$ supersymmetric gauge theories.
The first method is semiclassical and has lead to interesting
results, particularly for the $N=4$ theory. However, it
requires to deal with difficult mathematics, and 
cannot account, even in principle,
for the discontinuity of the spectrum at strong coupling.
The second method is non perturbative in nature, and has proved
to be very powerful and simple for the asymptotically free theories.
It provides a very easy way to compute the semiclassical spectrum,
and gives also the full answer at strong coupling.
The main challenge for the future will be to extend this method
to the conjectured self-dual field theories and try to get
here new insights on electric-magnetic duality.
\section*{Acknowledgements}
I am particularly grateful to the scientific directors of the '96 spring
workshop on string theory, gauge theory and quantum gravity
at ICTP, namely R. Dijkgraaf, R. Iengo, I. Klebanov,
K.S. Narain and S. Randjbar-Daemi, for inviting me to present
the material covered in these notes.


\begin{thebibliography}{99}
\bibitem{SWI}{N. Seiberg and E. Witten,
Nucl. Phys. B426 (1994) 19; B430 (1994) 485.}
%
\bibitem{SWII}{N. Seiberg and E. Witten,
Nucl. Phys. B431 (1994) 484.}
%
\bibitem{sen}{A. Sen, Phys. Lett. B329 (1994) 217.}
%
\bibitem{por}{M. Porrati, Phys. Lett. B377 (1996) 67.}
%
\bibitem{segal}{G. Segal and A. Selby, Comm. Math. Phys. 177 (1996) 775.}
%
\bibitem{FB}{F. Ferrari and A. Bilal, Nucl. Phys. B469 (1996) 387,
hep-th/9602082.}
%
\bibitem{BF}{A. Bilal and F. Ferrari, LPTENS-96/22, HUB-EP-96/11,
hep-th/9605101, to appear in Nucl. Phys. B480.}
%
\bibitem{BOG}{E.B. Bogomol'nyi, Sov. J. Nucl. Phys. 24 (1976) 449.}
%
\bibitem{ATH}{M. Atiyah and N. Hitchin, The geometry and dynamics of
magnetic monopoles, Princeton University press, Princeton, New Jersey.}
%
\bibitem{manton}{N.S. Manton, Phys. Lett. B110 (1982) 54.}
%
\bibitem{sut}{P.M. Sutcliffe, Phys. Lett. B357 (1995) 335\\
C.J. Houghton and P.M. Sutcliffe, Nucl. Phys. B464 (1996) 59\\
H. Braden and P. Sutcliffe, MS-96-016, UKC/IMS/96-64, hep-th/9610141.}
%
\bibitem{CAL}{C. Callias, Comm. Math. Phys. 62 (1978) 213\\
R. Bott and R. Seeley, Comm. Math. Phys. 62 (1978) 235.}
%
\bibitem{gau1}{J.P. Gauntlett, Nucl. Phys. B411 (1994) 443.}
%
\bibitem{AG}{L. Alvarez-Gaume and D.Z. Freedman, Comm. Math. Phys.
80 (1981) 443.}
%
\bibitem{ceder}{M. Cederwall, G. Ferretti, B.E.W. Nilsson and P. Salomonson,
Mod. Phys. Lett. A11 (1996) 367.}
%
\bibitem{gau2}{S. Sethi, M. Stern and E. Zaslow, Nucl. Phys. B457 (1995) 484\\
J.P. Gauntlett and J.A. Harvey, Nucl. Phys. B463 (1996) 287.}
%
\bibitem{witin}{E. Witten, Nucl Phys. B202 (1982) 253.}
%
\bibitem{ler}{A. Klemm, W. Lerche, P. Mayr, C. Vafa and N. Warner,
CERN-TH/96-95, HUTP-96/A014, USC-96/008, hep-th/9604034\\
A. Brandhuber and S. Stieberger, CERN-TH/96-263, NEIP-96/007,
hep-th/9610053.}
%
\bibitem{sei}{N. Seiberg, Phys. Rev. D49 (1994) 6857.}
%
\bibitem{ferrari}{F. Ferrari, LPTENS-96/54, hep-th/9609101.}
%
\bibitem{lind}{U. Lindstr\" om and M. Ro\c cek, Phys. Lett. B355
(1995) 492.}
%
\bibitem{BF2}{A. Bilal and F. Ferrari, Proceedings of the 
second international
Sakharov conference, Moscow, LPTENS-96/34, hep-th/9606111.}
%
\end{thebibliography}
\end{document}